\documentclass{elsart}

\usepackage{natbib}

 \usepackage{graphicx}

\usepackage{amssymb}

\begin{document}

\begin{frontmatter}



\title{Detecting X-ray QPOs in Active Galaxies}


\author[a1]{Simon Vaughan} and
\ead{sav2@star.le.ac.uk}
\author[a2]{Philip Uttley}
\address[a1]{X-Ray and Observational Astronomy Group, University of
   Leicester,  Leicester, LE1 7RH, UK}
\address[a2]{Laboratory for High Energy Astrophysics, Code 662, 
NASA Goddard Space Flight Center, Greenbelt Road, Greenbelt, MD 20771, USA}

\begin{abstract}
Since {\it EXOSAT} produced the first good quality X-ray
light curves of Seyfert galaxies there have been several claims of
quasi-periodic oscillations (QPOs). 
None of these have withstood repeated analyses and observations. 
We review some problems concerning the detection of QPOs.
\end{abstract}


\end{frontmatter}

\section{Introduction: Why look for QPOs in Active Galaxies?}
\label{sect:intro}

Active Galactic Nuclei (AGN) show strong, erratic brightness
fluctuations in the X-ray band (\citealt{law87}).  
At least for the Seyfert galaxies,
the power spectrum of these
variations is a power law-like continuum with a steep
slope, except in a few cases where the slope can be seen
to flatten at low frequencies (see e.g. \citealt{ump,mark}).
Processes with steep power spectra like this are known as
{\it red noise} (\citealt{press78}). 

The X-ray variations in Seyfert galaxies
bear more than a passing resemblance to those of X-ray binaries
(XRBs; e.g. \citealt{van95}).
In addition to a relatively featureless red noise continuum, 
XRBs often show QPOs in the
form of peaks  above the continuum (red noise)
power spectrum. By analogy with XRBs, AGN may also possess QPOs.
For example, assuming a $1/M_{BH}$ scaling of
frequencies we might expect to see analogues of the high frequency
QPOs of XRBs at $\sim 3 \times 10^{-3} (M_{BH}/
10^6 {\rm M_{\odot}})^{-1}$~Hz \citep{abr}.
The detection of periodic or quasi-periodic variations from
a Seyfert galaxy would be a major discovery and could lead to a
breakthrough in our understanding if the characteristic (peak)
frequency could be identified with some physically meaningful
frequency.  

\section{Textbook search methods}

One of the most popular tools for time series analysis
in astronomy is the {\it periodogram} \citep{sch98} which is simply the
modulus-squared of the discrete Fourier transform of the time series
(with an appropriate normalisation). This measures the amount of
variability {\it power} as a function of temporal frequency.
A concentration of power in a narrow range of frequencies 
would indicate a QPO\footnote{In the following we shall
denote any strictly or quasi-periodic oscillation with the
generic label ``QPO.''}. The notorious problem with the
periodogram of noise processes is that  
the ordinates are distributed like $\chi^{2}$ (with two degrees of
freedom) about the true 
spectrum \citep{pri81}. This means there are large fluctuations in the
periodogram that can
easily be mistaken for genuine peaks (periodic signals). 
Tests for periodic variations against a background of
white (flat spectrum) noise that take account of these intrinsic
sampling fluctuations are well established: see \citet{fis} or
section 6.1.4 of \citet{pri81}, and also \citet{lea}, \citet{van89}
and \citet{press96}. If the underlying source variability is
not white noise the problem is more difficult and these methods 
will yield unreliable results (\citealt{van89}; \citealt{ben}). 

\section{Possible sources of error}

\subsection{Systematic/instrumental effects}

With non-imaging detectors (e.g. {\it EXOSAT}, {\it Ginga}, {\it RXTE})
there is
always the possibility the light curve of a particular AGN
is contaminated by other bright X-ray sources within the field of
view. This possibility was realised in the case of NGC 6814 which had
been identified as the best candidate for a Seyfert galaxy with an
X-ray period
based on non-imaging {\it EXOSAT} observations (\citealt{mb}).  Subsequent
X-ray imaging observations made with {\it ROSAT} showed the periodic
oscillations  belonged to another source (V1432 Aquilae, a Cataclysmic
Variable) located $37$~arcmin from NGC 6814 (\citealt{mad93,staub}).
It is also possible for systematic, periodic changes in the detector
properties to artificially modulate the observed light curve. An
example is the {\it EXOSAT} data of NGC 5548. \citet{pap93a}
reported significant   periodicities  in this moderate luminosity
Seyfert but a reanalysis of the same data by \citet{tag} demonstrated
that one observation was affected by detector problems.
The likelihood of these problems is greatly reduced with the
current generation of low-background, imaging X-ray missions.

\subsection{Inappropriate null hypothesis: Assuming white noise}

Since the days of {\it EXOSAT} we have known that the
X-ray variations of Seyfert galaxies have an intrinsically
red power spectrum. This means the periodogram is not
flat in expectation (as for white noise) and is generally
not known {\it a priori}. As a result it is meaningless to
apply the standard `periodogram peak' tests
(\citealt{fis,lea,press96}). These all assume a null hypothesis of
independent (white), Gaussian noise. Thus a high
rejection probability for the null hypothesis obtained from any 
of these tests merely rejects
white noise as the sole origin of the variance.
It does not reject the more likely possibility that 
there are intrinsic red noise variations in the light curve
(i.e. non-white, but also not periodic variations).
Even a slight amount
of intrinsic red noise variance will raise the spectrum above the
white noise expectation and lead to spurious detections. 

For example,  \citet{iwa} discussed a possible periodicity
in IRAS $18325-5926$ from a long ($\sim 5$~day) {\it ASCA}
observation. The significance
of the candidate periodicity was estimated by testing against the
hypothesis of purely Poisson noise.
However, once the red noise variations intrinsic to the source were included in the
periodogram fitting the  candidate periodicity was no longer
significant at the $95\%$ level (\citealt{v04}).
Other Seyfert galaxies for which candidate QPOs were
reported without rigorously testing an origin in red noise
include 
RX J$0437.4-4711$ (\citealt{hm96}), 
MCG$-$6-30-15 (\citealt{lee00}), 
Mrk 766 (\citealt{bol}) and 
IRAS~$13224-3809$ (\citealt{pf}).

\subsection{Subtle power spectral effects}

\begin{figure}
\begin{center}
   \includegraphics[width=6 cm, angle=270]{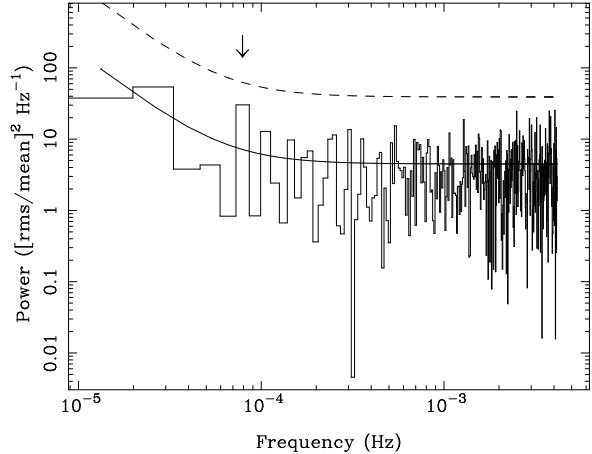}
\end{center}
\caption{
Periodogram of the 1986/060 {\it EXOSAT} observation of NGC 4151
measured from the background-subtracted $1-9$~keV ME data binned to $120$~s
(obtained through the {\it EXOSAT} data archive at HEASARC/GSFC).
The histogram shows the raw data and the solid curve marks the
Maximum Likelihood fit using a model comprising a power law (red
noise) plus constant (Poisson noise).
The dashed line represents a (minimum) $95\%$ confidence limit for the
detection of a QPO calculated as a factor $\gamma_{\epsilon}
= - \ln [ 1 - ( 1 - \epsilon )^{1/n} ]$ above the model, where $n$ is
the number of frequencies and $\epsilon$ is the `false alarm
probability'.
This quick calculation slightly underestimates the threshold
because it does not account for uncertainties in the modelling (see \citealt{v04}).
The arrow indicates the position of the candidate
QPO discussed by \citet{fiore}.
\label{fig:4151}
}
\end{figure}

There are a number of ways that one can account for
red noise in a light curve. But this is often 
not a trivial task and there are many possibilities
for overestimating the significance of candidate QPOs.
For example, \citet{fiore} reported candidate QPOs
in NGC 4151 based on three {\it EXOSAT} observations.
The authors claimed high ($>99\%$) significance 
peaks in the
periodograms after the data were ``pre-whitened''
by detrending the light curves using a parabola. However,
no account was made of the uncertainties
involved in this procedure, which would undoubtedly leave some
residual red noise component in the spectrum.
If the continuum power at low frequencies (where the QPOs occurred and 
where the red noise will be strongest) was underestimated by only
a small fraction this would drastically inflate the significance of
the candidate QPOs.
Figure~\ref{fig:4151} shows a re-analysis of one of these
datasets. Explicitly fitting the red noise and Poisson noise
components of the spectrum allows us to place more realistic
limits on the significances of any QPOs, which now fall
below the $95\%$ confidence detection threshold. 

The above example is one where the uncertainty in modelling
the aperiodic power spectrum (i.e. the intrinsic red noise and
instrumental Poisson noise) led to an artificially
enhanced periodicity significance. In general,
not accounting for the uncertainty in the null hypothesis model (the
continuum) often tends to overestimate the significance of any peaks in the residual
(``whitened'') periodogram.
For example, in their re-analysis of the {\it EXOSAT} observations of
NGC 5548, \citet{tag} found that in every case the significance of the
candidate QPO was
lower  than previously reported ($\le 95$\%) once due allowance was made
for the uncertainties in modelling the spectrum.

It is also possible to overestimate the significance 
of a periodogram peak when the periodogram is
oversampled. \citet{hlm}
identified possible QPOs in three Seyfert 1 galaxies
(Ton S180, RX J$0437.4-4711$, 1H 0419-577)
based on a uniform analysis of all the available long {\it EUVE}
observations. 
The significances of the periodicities were estimated
using a Monte Carlo procedure similar to that discussed by
\citet{ben}. 
However, the periodograms were oversampled
(computed at frequencies between, as well as at, the standard
Fourier frequencies). This increases the number of frequencies searched
and therefore also the chance of seeing a 
spurious signal  (\citealt{v04}).

Finally there are cases where genuinely significant
power spectral features have been detected
but were not caused by periodic oscillations. 
\citet{pap95} reported a broad bump  in the power
spectrum of NGC 4051 (based on {\it EXOSAT} observations).
However, recent observations with {\it ROSAT, RXTE} and {\it
  XMM-Newton} (\citealt{green, mch}) have provided much tighter
constraints on the power spectrum and have shown that
the underlying continuum spectrum of NGC 4051 breaks at a
similar frequency to that reported by \citet{pap95}.
The broad bump in the spectrum reported from the 
{\it EXOSAT} observation is therefore likely to be
due to the break in the continuum spectrum rather than an
additional QPO feature.

\section{Conclusions}

A major source of `false alarm' QPOs is the use of an
inappropriate continuum power spectrum.  
Arbitrarily assuming a white noise spectrum,
or red noise with an {\it ad hoc} slope, 
will alter the apparent strength (significance) of 
any peaks in the residual periodogram.
The continuum spectrum must be modelled explicitly.
Furthermore, one must account for uncertainties in
the continuum modelling.  Failure to include these may result in an
overly  optimistic significance calculation (\citealt{is, tag, v04}). 

The above arguments apply not only to
searching for peaks in the periodogram but also for searches based on
{\it epoch folding}, whereby one
bins the time series into phase bins at a
test period. At the correct period the periodic variations will
sum while any background noise will cancel out, thus revealing
the profile of the periodic pulsations. However, the background
noise will only cancel out if it is temporally independent, i.e.
white noise. Again, the presence of any underlying red noise
variations will produce unreliable results unless checked
with e.g.  Monte Carlo simulations\footnote{
A further word of caution is required
here.  For the correct application of Monte Carlo methods it is
crucial to use a random noise generator that  correctly produces the
statistical aspects of red noise and the observational sampling,
otherwise the significances may again be badly biased (\citealt{tim}).}
(\citealt{ben}).




\end{document}